\documentclass[letterpaper,twocolumn,10pt]{article}

\usepackage{usenix2019,epsfig}

\usepackage{listings}
\usepackage{todonotes}
\usepackage{multirow}
\usepackage{booktabs}
\usepackage{colortbl}
\usepackage{xcolor}
\usepackage{pifont}
\usepackage{textcomp} 
\usepackage{caption}
\usepackage{makecell}
\usepackage{paralist} 
\usepackage[hyphens]{url} 
\usepackage[hidelinks]{hyperref}	
\usepackage{tcolorbox}
\usepackage{footnote}
\usepackage{tablefootnote}
\usepackage{bigdelim}
\usepackage{listings}
\usepackage{mdframed}	
\usepackage{authblk}

\definecolor{KWColor}{rgb}{0.57,0.08,0.35}
\definecolor{CommentColor}{rgb}{0.133,0.545,0.133}
\definecolor{StringColor}{rgb}{0,0.126,0.941}

\newcommand{\TT}[1]{\lstinline[basicstyle=\tt,keywordstyle=,stringstyle=,breaklines=true]{#1}}

\newcommand{\component}[1]{\textsf{#1}}
\newcommand{\randfuzz}{\component{RandFuzz}}

\newcommand{\afl}{\component{AFL}}

\newcommand{\artist}{\component{ARTist}}
\newcommand{\troop}{\component{Troop}}

\newcommand{\uplink}{\component{Uplink}}

\newcommand{\tool}[1]{\textsf{#1}}
\newcommand{\frida}[0]{\tool{Frida}}

\definecolor{darkergreen}{rgb}{0.13, 0.55, 0.13}
\newcommand{\cmark}[0]{\textcolor{darkergreen}{\ding{51}}}
\newcommand{\xmark}[0]{\textcolor{red}{\ding{55}}}

\newcommand{\unknown}[0]{\textcolor{orange}{\textsf{?}}}

\newcommand{\circled}[1]{\raisebox{.5pt}{\textcircled{\raisebox{-.9pt} {#1}}}}

\newcommand{\privflaws}{11}

\definecolor{mygray}{gray}{0.85}
\definecolor{myblue}{rgb}{0.74, 0.83, 0.9}

\mdfdefinestyle{punchline}{roundcorner=20pt,backgroundcolor=lightgray}
\newenvironment{punchline}
{\begin{mdframed}[style=punchline]}
	{\end{mdframed}}

\begin{document}

\date{}

\title{Towards a Principled Approach for Dynamic Analysis of Android's Middleware}

\author[1]{Oliver Schranz}
\author[1]{Sebastian Weisgerber}
\author[2]{Erik Derr}
\author[1]{Michael Backes}
\author[1]{Sven Bugiel}
\affil[1]{CISPA Helmholtz Center for Information Security, Saarland Informatics Campus}
\affil[ ]{\textit{\{schranz, weisgerber, backes, bugiel\}@cispa.saarland}}
\affil[2]{University of Luxembourg}
\affil[ ]{\textit{derr@svv.lu}}

\maketitle

\subsection*{Abstract}

The Android middleware, in particular the so-called systemserver, is a crucial and central component to Android's security and robustness. To understand whether the systemserver provides the demanded security properties, it has to be thoroughly tested and analyzed. A dedicated line of research focuses exclusively on this task. While static analysis builds on established tools, dynamic testing approaches lack a common foundation, which prevents the community from comparing, reproducing, or even re-using existing results from related work. This raises questions about whether the underlying approach of any proposed solution is the only possible or optimal one, if it can be re-used as a building block for future analyses, or whether results generalize. 

\noindent
In this work, we argue that in order to steer away from incompatible custom toolchains and towards having comparable analyses with reproducible results, a more principled approach to dynamically analyzing the Android system is required. 
As an important first step in this direction, we propose a unified dynamic analysis platform that provides re-usable solutions for common challenges as the building blocks for future analyses and allows to compare different approaches under the same assumptions.

\section{Introduction}
\label{Introduction}

Android dominates the mobile market with a share of 71\%~\cite{marketshare}. The ecosystem consists of a multitude of customized Android versions
(ROMs) that are based on official releases of the Android Open Source Project (AOSP) maintained by Google. The rapid growth of AOSP's code base 
also affects one of Android's core components, the systemserver, an always-running, privileged process that implements the bulk of the Android application framework that provides core functionality to apps via a set of APIs, such as package management, activity service, or location service. Between Android 4.1 (released 2012) and Android 10 (released 2019) the
number of services has more than doubled from 65 to 166. As a consequence, the growing code complexity also increases the systemserver's attack surface by
adding more highly-privileged code and APIs.
Over the last years, research has identified new flaws and vulnerabilities of the systemserver that allow apps to elevate privileges~\cite{gong} or mount denial-of-service (DoS)
attacks due to, e.g., input validation errors~\cite{exhunter,bindercracker}, concurrency bugs~\cite{asv}, or inconsistent
permission enforcement~\cite{arcade, kratos, permupgrade, arf, acminer}.  Given its central role in the Android software stack, the
increasing amount of discovered problems emphasizes the need for thorough testing of the systemserver's robustness and security.

\noindent
Static and dynamic analysis have both been successfully applied to this problem domain because they come with complementary advantages and disadvantages. Static analysis provides better scalability and more formal guarantees by considering all possible execution traces, but suffers from an increased false positive rate because it also considers impossible code paths~\cite{undecidability}. Dynamic analysis avoids false positives by tying its results to actual execution traces, which, however, leads to scalability issues and requires a thorough code coverage of the test subject. Hybrid approaches exist where either static analysis pre-filters interesting candidates to restrict the amount of test subjects for dynamic analysis, or dynamic analysis is used to verify potential findings from static analysis.
Despite the shown efficacy of both kinds of analysis in certain cases, there is a major asymmetry in the literature because most static analyses on Android build upon a set of common base frameworks, such as Soot~\cite{soot} or Wala~\cite{wala}, which allows related work to compare against and build upon each other. Or in other words: there exists a principled approach to statically analyzing Android's middleware.
Dynamic analysis, in contrast, is lacking such a common foundation.
Instead, related work on dynamically analyzing Android's system services created a large set of mostly incompatible and highly specialized solutions. %
While they succeeded in implementing a multitude of analyses, such as vulnerability detection and permission mapping, their results are hard to compare or reproduce, and their solutions to shared challenges cannot be re-used as building blocks for future analyses. As a consequence, without a more principled approach to dynamically analyzing the Android system, it is unnecessarily hard to learn from those prior results, to identify more promising solutions and strategies, or to find and address shared challenges.
We believe that the driving factor for this discrepancy is the particular set of hurdles that comes with dynamically analyzing the systemserver. Its complex interdependencies with other Android core components prevent restarting or resetting the systemserver in isolation. This is exacerbated by its statefulness and asynchronous lifecycle design that make it hard to isolate execution traces and require frequent restarts and cleanups.
As a result, existing off-the-shelf components from related fields, such as fuzzers or instrumentation frameworks, are often not directly applicable, which leads to the creation of the custom toolchains that fragment this research area.

\noindent
In order to base these efforts on a common foundation,
we propose a more principled approach to systemserver dynamic analysis that favors reproducible results and inter-operable solutions that allow other researchers to build their systems on top of each other and make their results comparable. To make a first step, we propose a full-fledged dynamic analysis platform for Android's middleware that allows to combine and implement modular solutions to the challenges we identified from related work and removes the burden for future dynamic analyses of implementing the surrounding infrastructure and instead allows to focus on the actual analysis.
We showcase the benefits of our platform by implementing two use cases from related work, vulnerability detection and permission mapping, which we utilize to study and evaluate different strategies. 
In our comparative evaluation, we find that, in contrast to related work that heavily specializes on fine-grained vulnerability classes, our generic approach is already able to uncover flaws from multiple such classes without specialization.  
For permission mapping, we find that proven strategies, such as coverage-guided fuzzing, fail to deliver superior results in comparison to much simpler black-box alternatives.

\noindent In summary, we make the following contributions:

\begin{itshape}
\begin{compactenum}
	\item We thoroughly study related work to deduce a set of requirements for a shared foundation for dynamically analyzing Android's systemserver.
	\item We propose a common platform for implementing dynamic analyses on top of re-usable building blocks for the identified challenges to allow the community to evaluate, compare, and build upon related work.
	\item Using two analyses implemented on top of our platform, we show that principled evaluation is a valuable tool towards finding optimal strategies for specific analyses. 
	\item In order to facilitate independent research and evaluation, we open source our full platform and toolchain with currently 18 repositories in total (see Appendix~\ref{Apx:Tools}).
\end{compactenum}
\end{itshape}

\section{Challenges}
\label{Challenges}

As a first step towards a more principled approach, we need to understand the core challenges that are shared among related work.

\begin{figure}
	\includegraphics[scale=0.4]{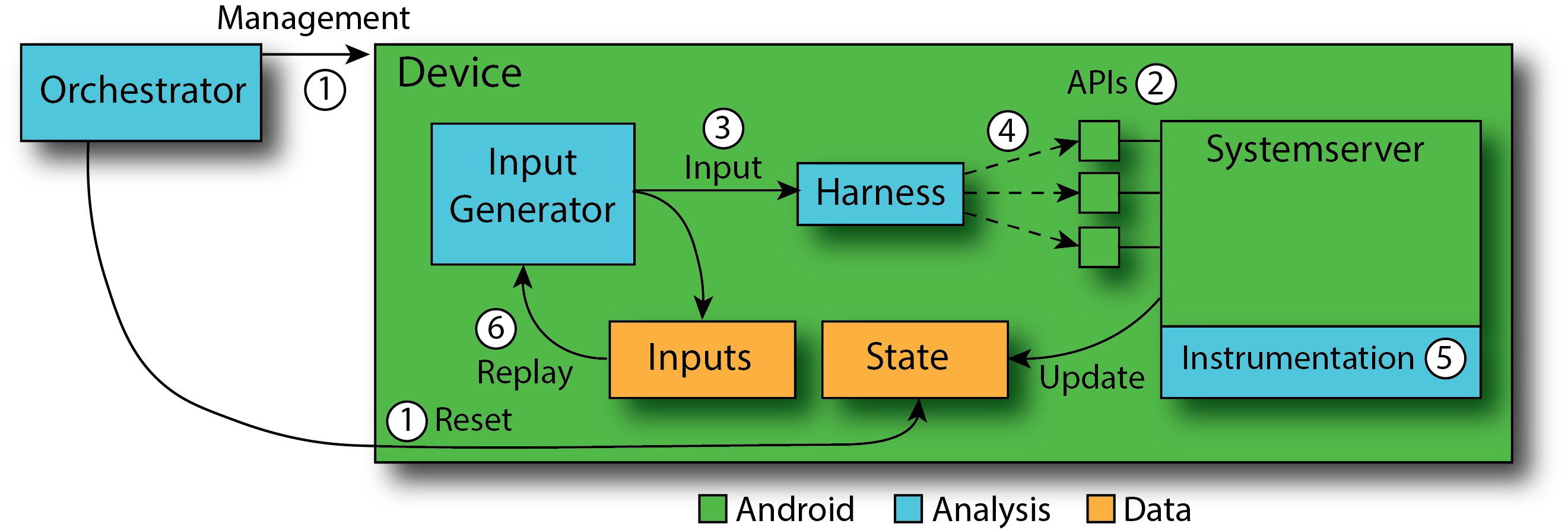}
	\caption{The typical structure of systemserver dynamic analysis solutions with commonly used components.}
	\label{fig:generic}
	\vspace*{-0.6cm}

\end{figure}

Figure~\ref{fig:generic} depicts the generalized structure of dynamic analyses of the systemserver that we derived from related work. Numbers \circled{1} to \circled{6} mark different key parts that each comes with a particular set of challenges that we discuss in the corresponding Sections~\ref{challenges:target} to~\ref{challenges:verification}. 
Table~\ref{tab:relatedwork} gives an overview of both academic and industry work and details their design decisions to which we will refer in the next sections. Furthermore, we provide an outlook on how each challenge is addressed by our platform.

\begin{table*}
	\scriptsize
	\centering
	\caption{Overview of existing systemserver dynamic analyses. }
		\resizebox{\textwidth}{!}{
		\begin{tabular}{cccccccccccc}
			\toprule
													& \multirow{2}{*}{\textbf{Analysis}}			& \multirow{2}{*}{\textbf{Devices}}	& \multirow{2}{*}{\textbf{Freshness}}	& \multirow{2}{*}{\textbf{Mapping}}							& \textbf{Input}	& \multirow{2}{*}{\textbf{Harness}}	& \multirow{2}{*}{\textbf{Coverage}}	& \textbf{Crash}						& \multirow{2}{*}{\textbf{Transformations}}							& \textbf{Auto-}	& \textbf{Exploit }				\\
													& & & & & \textbf{Generation} & & & \textbf{Detection} & & \textbf{Verification} & \textbf{Generation} \\
			\midrule

Chizpurfle (BB)~\cite{chizpurfle1, chizpurfle2}		& VD	& physical				& 	\xmark					& SM, RE				& own				& Java CLI tool		& \xmark					& LC, BD						& \xmark													& \xmark	& \xmark														\\
\rowcolor{mygray}
Chizpurfle (Evo)~\cite{chizpurfle1, chizpurfle2}	& VD	& physical				& 	\xmark					& SM, RE				& own				& Java CLI tool		& \cmark	& LC, BD						& coverage									& \xmark	& \xmark														\\

ExHunter~\cite{exhunter}							& VD	& physical				& 	\xmark					& SM, RE				& own				& App (Parcels)		& \xmark 				& LC, RD	& 	\xmark							& \cmark	& \cmark 				\\

\rowcolor{mygray}
ASVHunter*~\cite{asv}								& VF				& physical				& -						& SA							& only replay		& App				& \xmark					& \unknown												& manual: add logging to Binder		& \xmark	& \cmark								\\

Gong~\cite{gong}									& VD	& physical				& \xmark						& SM (native)					& own				& Binary (Binder)	&  \xmark				& \unknown								& \xmark								& \xmark					& \cmark								\\

\rowcolor{mygray}
Stowaway*~\cite{demystified}						& PE		& \unknown					& 	\xmark					& SM, RE				& Randoop~\cite{randoop}			& App				& \xmark					& 						-						& manual: log permissions	& \xmark & -	\\

BinderCracker (BB)~\cite{bindercracker}				& VD	& physical				& 	\xmark					& SM (native)	& own				& App				& \xmark					& \unknown												&  \xmark	& \xmark & \xmark \\

\rowcolor{mygray}
BinderCracker (Ctx*)~\cite{bindercracker}			& VD	& physical				& 	\xmark					& SM (native)	& own				& App				& \xmark					& \unknown												& manual: log binder transactions		& \xmark & \xmark\\

Buzzer~\cite{buzzer}								& VD	& \unknown					& 	\xmark					& SM (native)	& own				& App + native		& \xmark					& LC										& include hidden SDK APIs				& \xmark & \xmark\\

\rowcolor{mygray}
He~\cite{he}										& VD			& \unknown					& 	\xmark					& SP, IDA (static)& own				& native (\unknown)		& \xmark					& \unknown												& binder parcel interception			& \xmark & \cmark\\
FANS~\cite{fans}\tablefootnote{FANS explicitly targets native services only but shares many problems with systemserver dynamic analyses.}	& VD	& physical	& \xmark	& SP	& own	& Binary (Binder)	& \xmark	& LC,TS	& ASan~\cite{asan}	& \xmark	& \unknown \\

\midrule
\multicolumn{12}{c}{\textbf{Our Case Study Prototypes}}	\\
\midrule
\multirow{2}{*}{Vulnerability \& Bug Hunting}										& \multirow{2}{*}{VD}	& \multirow{2}{*}{emu}	& \multirow{2}{*}{per API}	& \multirow{2}{*}{SM (native), RE, SA}	& \afl~\cite{afl},Gong*~\cite{gong},	& \multirow{2}{*}{Binary (Binder)}, \multirow{2}{*}{Java CLI tool}	& \multirow{2}{*}{\cmark}	& \multirow{2}{*}{LC,PM,IN}	
														& \multirow{2}{*}{coverage, crash detection}	& \multirow{2}{*}{\cmark}	& \multirow{2}{*}{\cmark}	\\
& & & & & Chizpurfle~\cite{chizpurfle1,chizpurfle2} & & & & & \\

\multirow{2}{*}{Permission Mapping}									& \multirow{2}{*}{PE/VF}	& \multirow{2}{*}{emu}	& \multirow{2}{*}{per API}	& \multirow{2}{*}{SM (native), RE, SA}	& \afl~\cite{afl},Gong*~\cite{gong}							& \multirow{2}{*}{Binary (Binder), Java CLI tool}						& \multirow{2}{*}{\cmark}	& \multirow{2}{*}{LC,PM,IN}
														& \multirow{2}{*}{coverage}	& \multirow{2}{*}{\cmark}	& \multirow{2}{*}{\cmark}	\\ 
& & & & & Chizpurfle~\cite{chizpurfle1,chizpurfle2} & & & & & \\
			\bottomrule
		\end{tabular}
		}
	\caption*{\scriptsize
		\textbf{Legend:\\}
		\emph{Analysis}: \emph{VD}: vulnerability detection, \emph{VF}: verification, \emph{PE}: permission mapping \\
		\emph{Freshness}: How frequently device state is reset to avoid influencing the testing of another target.\\
		\emph{Mapping}: \emph{SM}: Service Manager, \emph{RE}: Reflection, \emph{SA}: static analysis, \emph{SP}: source parsing, \emph{IDA}: IDA Python plugin \\
		\emph{Coverage}: Whether coverage feedback is utilized.\\
		\emph{Crash Detection}: \emph{LC}: Logcat parsing, \emph{BD}: binder death notifications, \emph{RD}: app-based reboot detection, \emph{PM}: process monitoring, \emph{IN}: instrumentation, \emph{TS}L Tombstone monitoring

	}
	\label{tab:relatedwork}
	\vspace*{-0.6cm}

\end{table*}

\subsection{Target Instance Management}
\label{challenges:target}

Analyzing the systemserver comes with a particular set of challenges. Its complex interdependencies do not allow to run the systemserver independently of Android's software stack, hence separating executions or resetting state involves restarting or even resetting the whole operating system. Similar to related work that fuzzes, e.g., the Linux kernel~\cite{kAFL}, targeting Android system components requires explicit management of target instances (i.e., emulators and devices). 

To the best of our knowledge, related work in Table~\ref{tab:relatedwork} only utilizes physical Android devices for their evaluations (column \textbf{devices}), which is an indicator for why resetting the current state or isolating test cases was not implemented (column \textbf{freshness}). While Chizpurfle~\cite{chizpurfle1, chizpurfle2} explicitly targets vendor ROMs that often lack dedicated emulators, ASVHunter~\cite{asv}, Stowaway~\cite{demystified}, BinderCracker~\cite{bindercracker} and Buzzer~\cite{buzzer} require manual AOSP patches, which could have been deployed on emulators as well.

\begin{punchline}
	To cover different use cases, our platform supports both, physical and virtual
	devices, and exposes their configuration via APIs (e.g. number of instances, frequency of
	refreshs).	
\end{punchline}

\pagebreak

\subsection{Attack Surface Mapping}
\label{challenges:mapping}

The concrete attack surface exposed by the middleware completely depends on the currently running ROM. There are at least two dimensions to consider: First, in addition to AOSPs base set of system services and APIs that are required for Android to work properly (enforced by Google through their Compatibility Test Suite~\cite{cts}), vendors further extend Android with new services and APIs, which led to an increase in flaws and vulnerabilities~\cite{vendorcustom,drivers} in the past. Second, the number of system services of AOSP increases steadily with the introduction of new major Android releases. For this reason, dynamic analyses of the middleware requires a mapping of which APIs are accessible for the ROM under test to decide which APIs to target.
Depending on the concrete tool, the API methods are either identified by their high-level Java method signatures or their low-level transaction IDs used in native code. 

\begin{table}
\caption{Categorization of different API surface mapping approaches in terms of the used analysis method and resulting API description.}
\resizebox{\linewidth}{!}{
\label{tab:mapping}
	\begin{tabular}{c|cc}
		\toprule
									& \multicolumn{2}{c}{Method identifiers}	\\
									&	Transaction IDs						& Method signatures							\\
		\midrule

		\multirow{2}{*}{dynamic}	& native service manager	& Java service manager + reflection \\
									& \cite{gong, bindercracker, buzzer}	&	\cite{chizpurfle1, chizpurfle2, exhunter, demystified}						\\
		\multirow{2}{*}{static}		& source code parsing					& class hierarchy analysis			\\
									&	\cite{he,fans}							&	\cite{asv}							\\
		\bottomrule
\end{tabular}
}

\end{table}

The row \textbf{Mapping} in Table~\ref{tab:relatedwork} lists the different approaches to API mapping by existing work.
Table~\ref{tab:mapping} further categorizes these approaches in two dimensions: the collected method identifier and the required kind of analysis to obtain these identifiers.

\noindent
\textbf{Dynamic approaches} can list all available APIs at runtime by talking to a central service manager. While their full signatures are obtained via Java reflection, the corresponding transaction IDs are obtained using native code.

\noindent
We currently see two \textbf{static approaches} in the literature. First, parsing the AOSP code for transaction IDs restricts the whole toolchain to open source ROMs but can directly relate the low-level transaction IDs to high-level method signatures. Second, using static analysis to identify all  systemserver APIs is applicable to arbitrary ROMs since it operates on the bytecode but is prone to generate false positives and false negatives due to well-studied restrictions of static analysis in this domain~\cite{pscout,axplorer,arcade}. 

\begin{punchline}
	We provide both mappings, high-level Java signatures and low-level transaction codes, to support tools on the Java and native layer.
\end{punchline}

\subsection{Input Generation}
\label{challenges:input}

Input generators are an integral component of dynamic analyses because they drive the actual executions. 
However, different kinds of dynamic analyses require particular types of input generators.  
We discuss these requirements for each of the three types of dynamic analysis from column \textbf{Analysis} of Table~\ref{tab:relatedwork}:

\noindent
\textbf{1. Vulnerability detection} is the largest group with nine different solutions from existing work. All of these systems rely on custom fuzzers to generate crashes, from denial-of-service attacks to severe low-level memory corruption errors. We make two observations from studying related work. First, most fuzzers are created for exactly this use case of uncovering unknown bugs and flaws, hence choosing this type of input generation for vulnerability detection on Android is intuitive because of a long-standing history of successes~\cite{aflhof,hongghof,ossfuzzhof}. Second, as depicted in column \textbf{Input} in Table~\ref{tab:relatedwork}, all vulnerability detection analyses ship their own custom fuzzers instead of re-using state-of-the-art tools. This is counter-intuitive because the amount of discovered bugs typically increases with the depth and coverage that is achievable by the fuzzer and there is a dedicated line of work solely focused on improving general purpose fuzzers~\cite{fuzzeval,vuzzer,afl,directedfuzzing,angora}.

\noindent
\textbf{2. Dynamic verification} of static analysis results, as implemented by ASVHunter, utilizes static analysis to produce a set of candidate vulnerabilities that are subsequently verified dynamically. Static analysis is often prone to generating false positives, because it resorts to over-approximation when in doubt\footnote{Cf. the asynchronous \textit{Handler} pattern in the systemserver~\cite{axplorer, arf}.}, so a dynamic complement is created to test whether these candidate inputs are actually triggering the expected behavior. In contrast to fuzzers, there is no need for massively generating inputs as we are working with a known set of candidate inputs from static analysis.

\noindent
\textbf{3. Permission mapping}, as done by Stowaway, focuses on triggering as many permission checks as possible to be able to enumerate and map them. 
Their approach of using a test generator is comparable to employing a fuzzer. However, while vulnerability detection aims for \emph{deep} coverage of the target under test, permission checks are often placed at the beginning of API methods (\emph{fail early} principle)~\cite{arcade}, so thorough code coverage is not required. 

\begin{punchline}

To satisfy different analyses' input generation requirements, we integrated multiple kinds of input generators into our platform as re-usable building blocks (see Section~\ref{integratedfuzzers}) and provide APIs to integrate more.
\end{punchline}

\subsection{Target Communication \& Harnesses}
\label{challenges:harness}

In contrast to other domains, such as binary testing where we directly feed input to targets, step \circled{4} in Figure~\ref{fig:generic} illustrates that in the systemserver analysis scenario we have to resort to techniques from API testing to combine input generators with test harnesses. These harnesses have to match inputs to the API's required format and are responsible to bridge the process boundary between generators and targets by transparently forwarding all generated inputs. 
We distinguish between four different techniques:

\noindent
	\textbf{1. Native Binder Interfaces}, as used by Gong, Buzzer, He and FANS, allow to immediately feed marshalled parameter structures to the systemserver. While this approach provides superior performance, the semantic insight is rather low because without auxiliary information the types and structure of the target API's parameters are unknown.

	\noindent
	\textbf{2. Android SDK Managers} provide app developers with stable APIs while handling communication with the corresponding system services in the background. While this abstraction allows to provide Java objects as inputs, the APIs exposed by the SDK do not match those of the corresponding system services (number of parameters, types) and additionally include client-side error handling logic. Therefore, the SDK API does not reflect the actual attack surface of the underlying system services and should thus be avoided.

	\noindent
	\textbf{3. Reflection-based approaches}, as seen in Chizpurfle, BinderCracker, Stowaway, and ExHunter, also operate on the semantic level of Java objects but come with an increased complexity and overhead. However, they bypass the abovementioned SDK checks and allow to directly talk to the system services while transparently handling the inter-proces scommunication (IPC).

	\noindent
	\textbf{4. Intra-Process Communication} with the systemserver requires to run the input generator in the target's process space. It is the fastest alternative since it completely skips IPC calls, but it comes with robustness issues (generator runs in target process) and might incur an increased false positive rate because sanity checks are skipped.

\begin{punchline}
	
	Since related work operates on the native \textit{and} the Java layer, we implement harnesses for our integrated input generators to support both layers. 
	
\end{punchline}

\subsection{Instrumentation \& Introspection}
\label{challenges:instrumentation}

Dynamic analysis is typically supported by instrumentation or introspection tools, e.g., to provide coverage feedback, detect crashes, or patch out obstacles. 
Column \textbf{transformations} in Table~\ref{tab:relatedwork} gives an overview of instrumentation approaches used by related work. Analyses focusing on detecting vulnerabilities and other misbehavior utilize different techniques to detect the underlying crashes: log parsing, Binder death notifications~\cite{binderdeath}, monitoring the liveness of the systemserver and other system processes, and detecting device reboots. However, there are also instances of manually patching AOSP to include more verbose logging or allow mutation of inputs directly in the target process (ASV, BinderCracker, He), and utilizing the ASan support of more recent AOSP versions~\cite{asan} (FANS). Furthermore, Stowaway records permission checks and Buzzer re-compiles the SDK to include hidden APIs. In contrast, Chizpurfle follows a different path by utilizing the \frida~\cite{frida} hooking framework to implement coverage feedback. 

\begin{punchline}

	Our platform provides a set of re-usable introspection modules that cover a large set of requirements from related work, and additionally integrates with an instrumentation framework for more specialized requirements.
\end{punchline}

\subsection{Verification of Results}
\label{challenges:verification}

In real-world deployments of dynamic analyses, false positives cannot be avoided completely, e.g., due to background activity of other (system) components. 

In order to pre-filter results for human analysts or even achieve full autonomy~\cite{angr,cgc}, a set of solutions has been proposed to programmatically deal with these problems.
Typical examples are automated confirmations for crashes found by fuzzers by systematically replaying previous inputs and applying test case minification (e.g., AFL~\cite{afl}).

As depicted in the column \textbf{Auto-Verification} in Table~\ref{tab:relatedwork}, most related work describes their result verification process as \textit{manual}, with two notable exceptions. ExHunter fully automates the process of finding so-called \textsf{UncaughtException}~\cite{exhunter} denial-of-service attacks against the systemserver by detecting device reboots and replaying the offending input to verify if the system reboots again, thereby even creating proof-of-concept exploits for these inputs. Similarly, the dynamic part of ASVHunter only cares about verifying potential denial-of-service flaws by triggering the candidate input and observing the result.

\begin{punchline}

We provide out-of-the-box support for reproducing interesting behavior by persisting all generated inputs and automatically replaying them on fresh Android instances.

\end{punchline}

\section{Platform Architecture}
\label{Design}
Given the outlined challenges and the (partial) solutions provided by related work, we base the design of our dynamic analysis platform on these observations and show how to overcome the corresponding challenges in a modular and reusable way. 
This allows us to directly compare different building blocks and their strategies. We can then answer questions, such as whether the approach used by related work was the \textit{best-possible} one, or if re-using off-the-shelf components like \afl{}, which gave outstanding results for binary analysis, will provide similar results in the domain of systemserver analysis. 

\subsection{Overview}
\label{ArchitectureOverview}

\begin{figure}
	
	\includegraphics[width=\columnwidth]{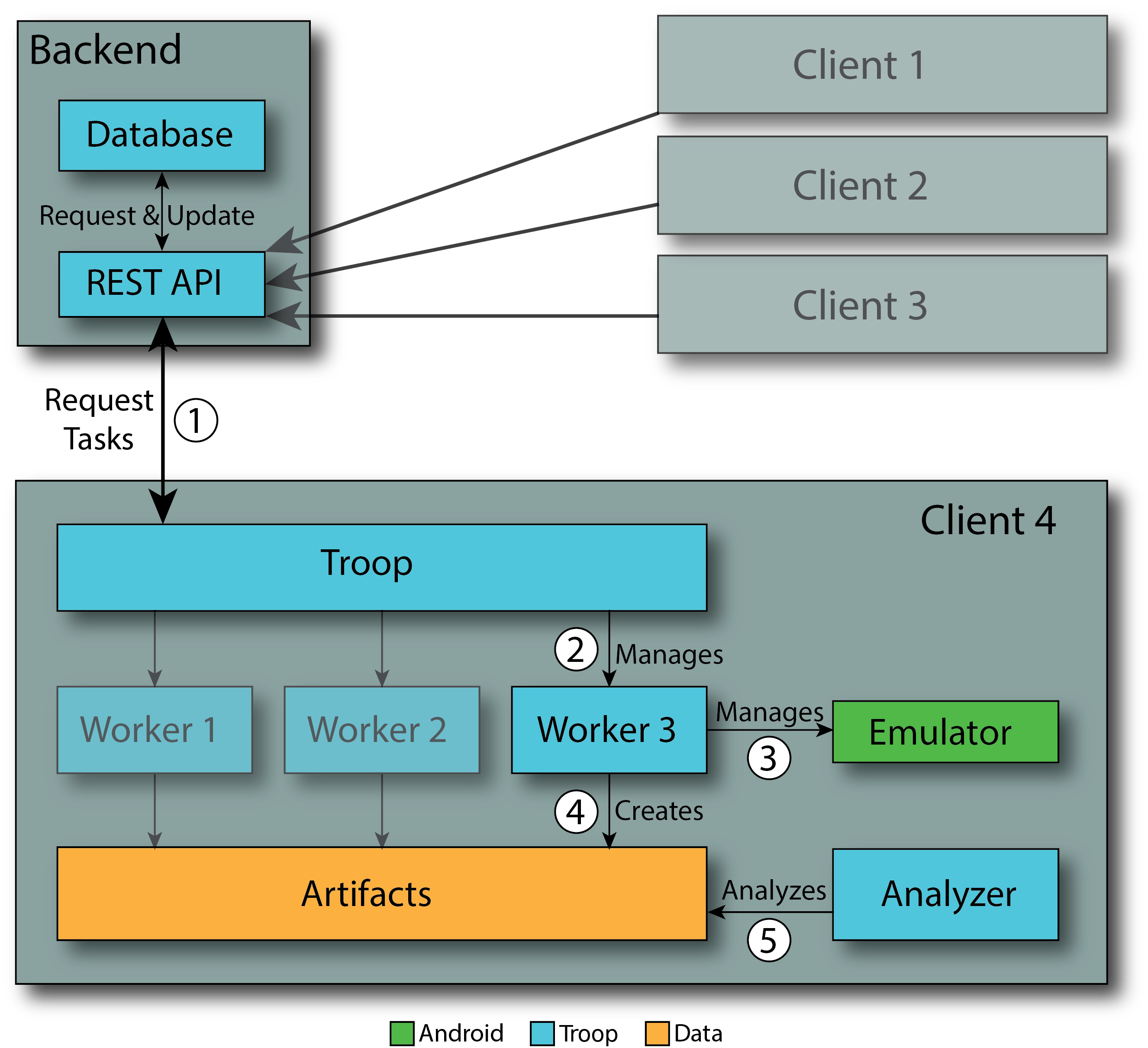}
	\caption{Overview of the platform's architecture.}
	\label{fig:overview}
\end{figure}

Figure \ref{fig:overview} gives an overview of our platform.
The clients request tasks \circled{1} (in this case sets of APIs to be analyzed) from a backend component and distribute them among their workers. 
An arbitrary number of clients running on different machines with varying amounts of workers attached to them are supported so that computational resources can be added and removed easily. 
The client's core is the manager component called \troop{} that distributes the work to locally running worker processes and takes care of monitoring all workers and exposing information about their current state \circled{2}. 
In our current design, each worker is responsible for exactly one (emulated or physical) device instance where it performs the analysis task at hand \circled{3} and creates corresponding artifacts and results \circled{4}. Additionally, a result analyzer component can interpret the worker's artifacts to detect findings, compute success rates, and provide further insights \circled{5}. 
The whole system is specifically designed for modularity because the components are loosely coupled, and for a high degree of automation to ensure scalability.

\subsection{\troop}
\label{Troop}

\troop{} is the central  entity that runs on each client and takes care of bootstrapping the whole process. It functions as a central observer that spawns all worker processes, continuously feeds them with new work from the backend, monitors their execution, and takes care of other issues, such as logging and management of generated test artifacts. 
It exposes the live data collected from its workers via a simple flask-based~\cite{Flask} webserver that describes the workers' current states in both a human-readable HTML format and a machine-readable JSON dump. 
The management code is decoupled from the actual worker implementation that is used for the current evaluation, so \troop{} operates independently of whether its workers are fuzzing their targets, confirming bug candidates, or performing any other analysis task.

\subsection{Result Analyzer}
\label{Analyzer}

Our framework ships with a set of result analysis components that can be used and extended to reason about the current state of dynamic analysis evaluations and their results. Our result analyzer modules scan through the artifacts generated by workers to produce summaries of the results. It can be invoked from the command line to get an idea about the current state of the evaluation but it is also used by \troop{} and the workers themselves for decision making. Typically, during the design of a new evaluation, interesting observations and bugs are encoded as new analyzer commands to track them as the system evolves. All results presented in Section~\ref{CaseStudy} are based on those result analyzer outputs.

\subsection{Backend}
\label{Backend}

The backend is a small web server that exposes the list of tasks stored in its database via a REST API. The most common cases are clients requesting new tasks for their workers to test. The backend remembers tasks currently under test, which ones have been finished, and a list of open tasks, thereby synchronizing all existing clients and workers to avoid duplicate work.

\subsection{Workers}
\label{Workers}

Workers implement the central logic of an evaluation because they directly perform the required tasks. After being started by \troop, they continuously consume tasks from their input queues and, depending on the kind of task, execute their functionality. There are no restrictions imposed by the platform on what kind of analysis or testing can be applied, so whether the worker runs a fuzzer, replays inputs, or uses a completely different input generator, is up to the concrete evaluation at hand. 
Workers alleviate the test case isolation and statefulness problems from Section~\ref{challenges:target} by launching \textit{fresh} instances of prepared emulators and managing their lifecycle including setup and teardown, so dynamic analyses only need to decide \textit{how often} to reset the testing environment, e.g., for each new tested API or after a fixed time budget is exhausted. 

\subsubsection{Workflow}

\begin{figure}
	
	\includegraphics[width=\columnwidth]{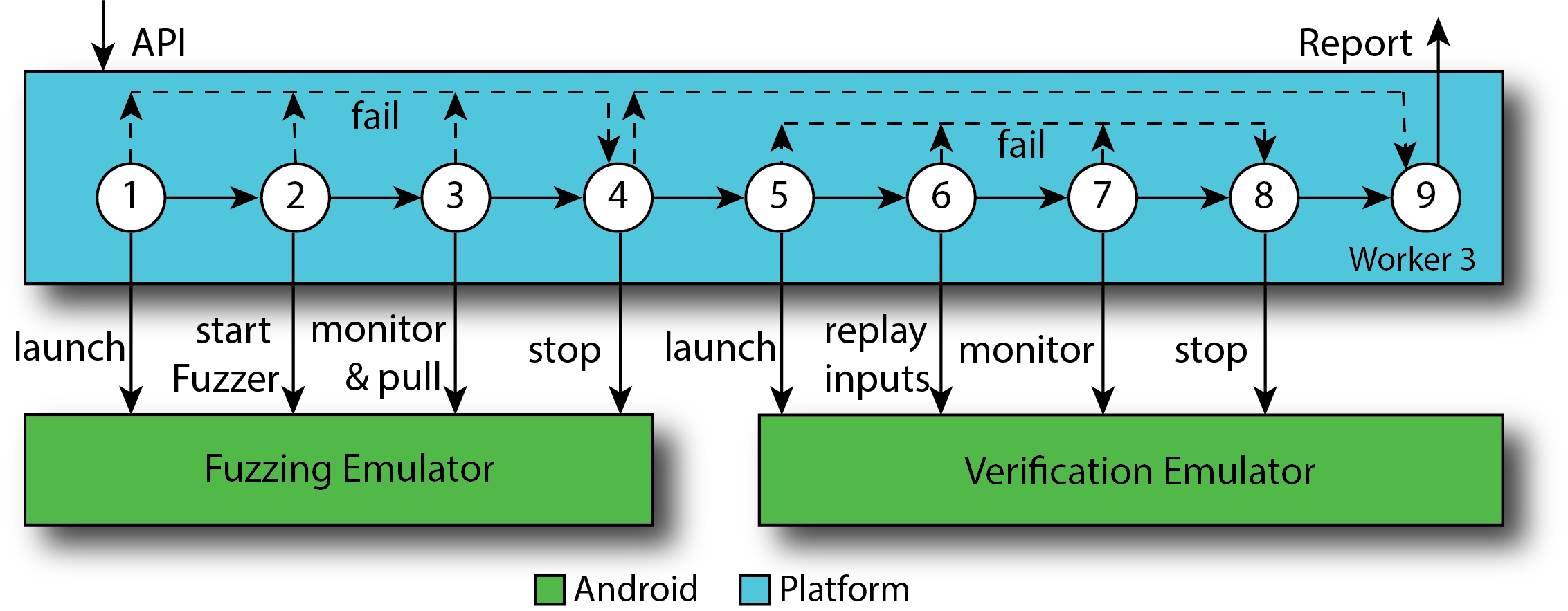}
	\caption{Worker processing an API task.}
	\label{fig:fuzzworker}
\end{figure}

Figure~\ref{fig:fuzzworker} depicts the typical worker pipeline for an input generator-driven analysis, in this example a fuzzer.
The worker starts by launching a \textit{fresh} emulator instance \circled{1}. Next, the input generator is launched to continuously send inputs to the current API\footnote{For the sake of simplicity we analyze one API at a time for now.} on the emulator \circled{2} while the worker regularly checks the status of all involved parties \circled{3}, e.g., if the input generator is still running and the system is behaving as expected. Furthermore, it keeps pulling snapshots of the current state of all platform components running on the emulator. After a predefined timeout, or as soon as we detect a potential finding, a snapshot is pulled and the whole emulator is shut down \circled{4}. 
In the next step, we try to verify the potential problem on vanilla emulators \circled{5}. 
Since our goal is to confirm whether the collected input list reliably reproduces the observed behavior, we minimize changes to this emulator by only using non-intrusive monitors (e.g., off-device monitors).
The list of previously generated inputs is replayed~\circled{6} and the system observes the emulator to decide whether the behavior is reproducible or a false positive~\circled{7}. 
Our verification workflow immediately outputs test cases that reproduce the finding. Finally, emulators are closed \circled{8}, files are backed-up, a report is written \circled{9}, and the worker starts all over with the next API.

\subsubsection{Monitoring}
\label{monitoring}

In order to fulfill the requirements from Section~\ref{challenges:instrumentation}, we employ a set of generic introspection modules for monitoring the system's current state to, e.g., detect crashes or collect coverage information. These modules check for interesting behavior as the result of the currently executed dynamic analysis and confirm that all framework components are still alive. To this end, we use four different techniques:

\noindent
\textbf{1. Logcat} is constantly parsed for crashes, stack traces, and other markers of unwanted behavior.

\noindent
\textbf{2. Checking the list of running processes} ensures we see crashes of critical Android services and platform components.

\noindent
\textbf{3. Uncaught exception handlers} for all vital systemserver threads are injected via instrumentation.

\noindent
\textbf{4. Coverage tracking} is implemented by instrumenting target methods to compute basic block and edge coverage\footnote{Implementing other coverage metrics discussed in the literature (e.g., path sensitive~\cite{collafl}) is also straightforward.}. 

\subsection{Integrated Input Generators}
\label{integratedfuzzers}

In Section~\ref{challenges:input} we learned that most dynamic analyses from related work utilize custom fuzzers tailored to their particular case. %
This not only leads to duplicate work but also misses out on utilizing improvements from the very active community that formed around advancing and optimizing general-purpose fuzzers (see Section~\ref{challenges:input}). 

Therefore, we decided to integrate a collection of different fuzzers that can be utilized by dynamic analyses, effectively freeing future work from re-inventing the wheel over and over again. This further allows us to compare their performance in different dynamic analysis scenarios to be able to pick the best-fitted fuzzer. 
In the following, we will discuss the three %
different fuzzers and how they are integrated into our platform.

\subsubsection{Chizpurfle}
\label{integratedfuzzers/chizpurfle}
We decided to integrate Chizpurfle~\cite{chizpurfle1,chizpurfle2} into our platform because it not only implements a single specialized fuzzer but intends to be a generic boilerplate for experimenting with different fuzzing techniques, such as black box or evolutionary algorithms. Since it already targets systemserver APIs, we do not need a harness and can directly instruct it to test particular APIs.

\paragraph{Coverage Channel.}
Chizpurfle utilizes the \frida~framework to trace executed code at runtime and provide coverage feedback to the fuzzer. However, we decided to use a more lightweight instrumentation framework instead for two reasons. First, the authors report an overhead of 1291\% introduced by \frida~\cite{chizpurfle1, chizpurfle2}, which tremendously slows down the fuzzer. Second, in our experiments, we could not make Chizpurfle's \frida~code run on x64 emulators, hence running the original version would restrict us to real ARM devices in contrast to a large number of emulators that we can run in parallel (see evaluations in Section~\ref{CaseStudy}). 
Since our coverage monitor from Section~\ref{monitoring} is implemented using the \artist~\cite{artist} instrumentation framework, we extend our \artist~module by additionally opening a local socket that the fuzzer can connect to and re-implement Chizpurfle's protocol to establish compatibility. The feedback channel is then used to transparently send the collected coverage back to the fuzzer in the expected format.

\subsubsection{RandFuzz (Gong)}
\label{case-studies/randfuzz}

With Chizpurfle as an example of a feedback-directed greybox fuzzer operating on the Java layer, we decided to also integrate a low-level fuzzer that uses the APIs transaction IDs instead. Under the name \randfuzz, we chose to re-implement Gong's approach of a random black box fuzzer based on the information from~\cite{gong}. 

\paragraph{Implementation \& Integration.}
From the available information, we implemented \randfuzz~to first list all accessible services by querying the native service manager (see Section~\ref{challenges:mapping}). \randfuzz~will then generate random payloads and send them to the targeted APIs. The fuzzer is executed by providing the target API's service names and transaction IDs.

\subsubsection{American Fuzzy Lop (AFL)}
\label{case-studies/afl}

In Section~\ref{challenges:input} we learned that, with the exception of Stowaway, no related work makes use of existing input generators but instead creates their own fuzzer from scratch. In order to showcase how to utilize state-of-the-art tools for analyzing the systemserver, we integrate the well-known American Fuzzy Lop (\afl~\cite{afl}) into the platform.

\paragraph{Bridging the Process Gap.}

Since the systemserver exposes multiple APIs from a remote process (see Section~\ref{challenges:harness}), we need to bridge the process gap to forward inputs to targeted APIs.
In our case, the fuzzing harness that we implemented takes the target service, transaction ID and input data, and transmits this information to the correct systemserver API via IPC. The forkserver that \afl{} usually employs to have multiple instances of its target running in parallel, in this case its harness, is deactivated for our purposes. All running harness instances are still backed by a single systemserver so parallel executions are not improving performance but they increase complexity in terms of isolating executions of generated inputs.

\paragraph{Input Generation.}

\afl{} typically outputs byte buffers that need to be structured according to the input format of the subject under test. To that end, we translate the input to primitive (Java) values that can be easily marshalled and sent across the process boundary. 
We abstain from leveraging \afl{} for fuzzing APIs with complex types for now, as this requires careful object modeling in order to reach the target API and to not extensively fuzz the sanity checks.

\paragraph{Coverage Channel.}
By default, \afl{} compiles a binary using a customized LLVM to insert feedback instructions into the code such that coverage information is written to a shared feedback buffer. This raises two concerns: First, we need to share a memory buffer between the systemserver process and \afl{}. Second, \afl{} cannot compile the systemserver, hence, we need another way to write feedback to the shared buffer. 

We solve the shared buffer problem by introducing an on-device component called \uplink{} that allocates \afl's shared buffer and shares the corresponding file descriptor with the systemserver and \afl. 
It's first client is the Java-based library that \artist{} injects into the systemserver to ensure we obtain a file descriptor for writing coverage feedback. The second client is \afl{}, which receives coverage feedback from the shared buffer. 

Concerning the second problem, we note that LLVM-based instrumentation is not always feasible (e.g., code unavailable, complicated dependencies) and therefore \afl{} supports a mode of operation where it skips the LLVM-based instrumentation process and just relies on an external tool to write feedback to the shared buffer.
Similarly to our integration of Chizpurfle~(see Section~\ref{integratedfuzzers/chizpurfle}), we can use the coverage monitoring \artist~module to push the collected coverage information back to the fuzzer, this time by writing it to the shared buffer obtained from \uplink.

\subsection{Device Setup}

\begin{figure}
	\includegraphics[width=\columnwidth]{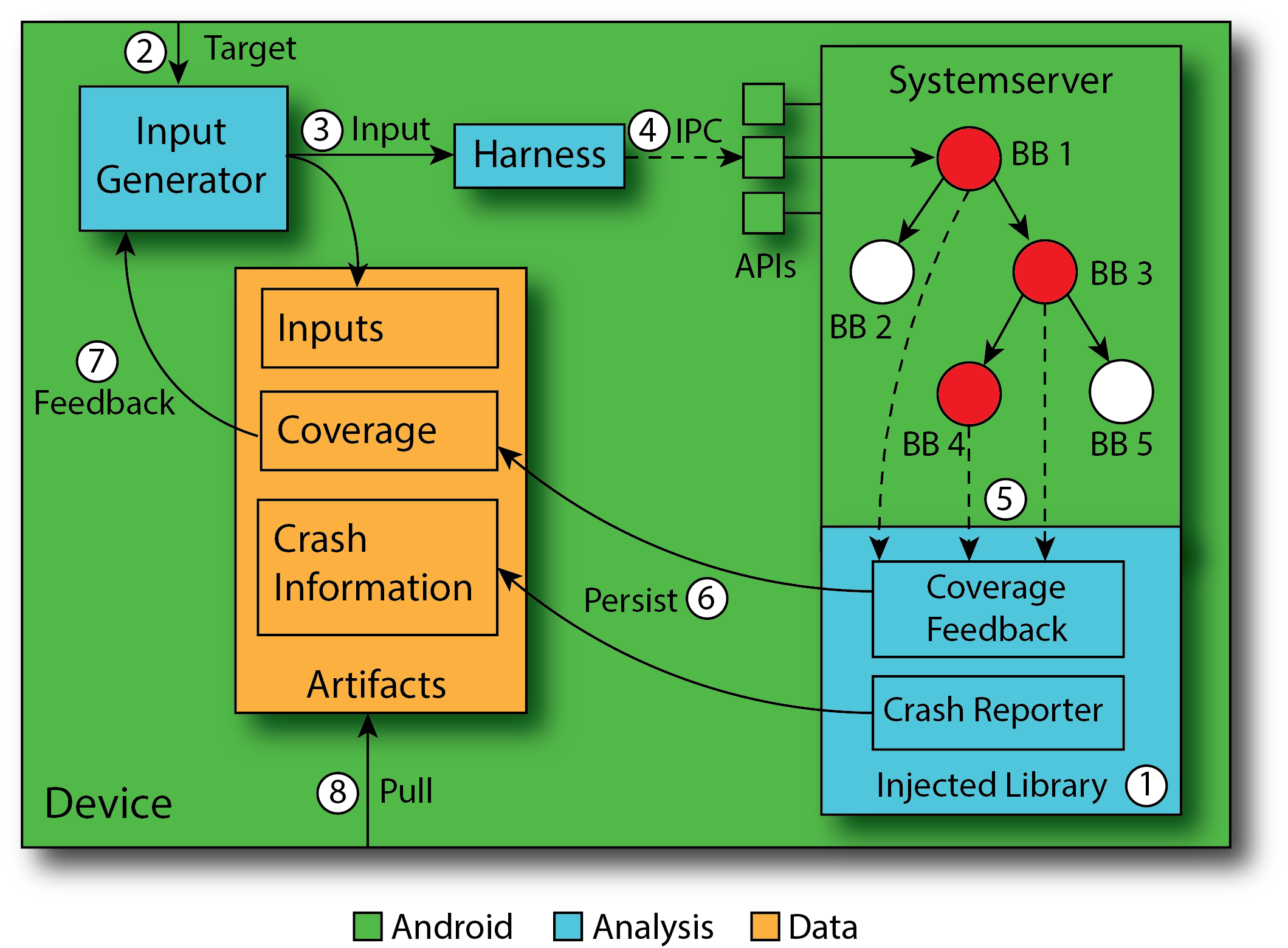}
	\vspace*{-0.5cm}
	\caption{The on-device setup from generating inputs to persisting and feeding back artifacts.}
	\label{fig:Emulator}
\end{figure}

Combining all aforementioned tools and modules, we obtain the device setup depicted in Figure \ref{fig:Emulator}. At boot time, \artist{} executes our modules that instrument the systemserver for crash detection and coverage feedback~\circled{1}. %
After Android booted successfully, the chosen input generator is started with the current API target~\circled{2} and generates inputs~\circled{3} that are transparently sent to the systemserver via Binder IPC~\circled{4}. On the server side, the API is executed and the code injected by our coverage module notifies the injected Java library about each basic block that is executed~\circled{5}.
Coverage and crash information is persisted~\circled{6} and coverage feedback is provided back to the input generator~\circled{7}.
The generation of new inputs is guided by the feedback received from the coverage module, hence executions discovering new basic blocks directly influence the generation of new inputs. 
Off-device the responsible worker periodically checks the health of all involved components (e.g., systemserver, input generator, monitors), collects \textsf{logcat} messages, pulls system tombstones (Android crash dumps), and persists the collected artifacts~\circled{8}.

\section{Case Studies}
\label{CaseStudy}

In this section, we study how to use our platform to approach dynamic analysis tasks from related work in a more principled way.
The two case studies we conduct show how our platform can be utilized to explore different design decisions and  strategies, which we exemplify by comparing the impact of input generation strategies, such as the choice of fuzzers and whether to utilize coverage feedback. Our results indicate that there are still lots of open questions in order to find \textit{optimal} components for the different dynamic analyses.

\subsection{Vulnerability \& Bug Hunting}

Vulnerability and bug hunting is often perceived as a \textit{dark art} driven by intuition and gut feeling. As evident from Table~\ref{tab:relatedwork}, there has been a multitude of approaches for hunting bugs in the systemserver and other middleware components. While they succeeded in uncovering different kinds of flaws, they rely on customized toolchains that are mostly incompatible and incomparable. This raises questions such as whether the implemented strategies are the most beneficial ones, or where we can still improve in those areas. While these questions currently cannot be answered for the Android domain, there has been a lot of work in related areas recently that focuses on properly comparing and evaluating dynamic analysis approaches and their central components (e.g., input generators) to create benchmarks and evaluation criteria~\cite{fuzzeval,angr,lava}. For dynamically analyzing the systemserver, however, we are lacking such a principled approach that places related work and their results on firm ground. 
We conduct a case study on vulnerability and bug hunting on top of our common platform to answer questions such as whether coverage-guided approaches outperform black-box approaches and how the required instrumentation impacts performance.

\subsubsection{Design}

We replicate the scenario of related work from Table~\ref{tab:relatedwork} by designing our case study to search for crashes in the systemserver. %
To be able to replace and compare components easily, we fully utilize all existing building blocks: The three integrated fuzzers (Section~\ref{integratedfuzzers}) are generating inputs, the monitors (Section~\ref{monitoring}) feed back coverage and crash information, and \troop{} (Section~\ref{Troop}) parallelizes the testing across multiple emulators.
As soon as a monitor detects a crash and all information is pulled, we use the auto-verification support to replay the collected inputs on a clean vanilla emulator. The input replaying component together with the list of inputs thereby form a proof-of-concept exploit. This platform-driven approach for the first time allows us to conduct a comparative evaluation to assess the impact of different design decisions.

\subsubsection{Comparative Evaluation}

The platform's modular design allows us to evaluate different components under the same conditions while leaving all other components fixed between tests. Since most related work employs fuzzers to drive their concrete executions, we create seven different setups based on our integrated fuzzers to study their impact in terms of performance and coverage.

\noindent
\paragraph{Setup.}
\label{casestudies:setup}

All experiments are executed with 30 parallel workers analyzing each target API for 5 minutes. The emulators run a custom AOSP 7.1 image that includes additional tooling required for the evaluation.
We compute the target APIs for the experiments by first enumerating all methods from the 89 Java-backed system services. Out of the 2,090 interface-method-pairs, we were able to map 2,005 APIs to concrete implementations and generated coverage information for 1,867 (89\%) of them, where we only consider those 1,637 (78\%) with a non-empty list of parameters.
Subsequently we divide our experiments into two groups based on the APIs they are targeting.
The first group contains all 433 target APIs that are currently supported by \afl{} and \randfuzz{} because all their parameters have primitive or string types and we have binder transaction IDs available, hence we call this group the \emph{primitive APIs}.
The second group, which contains all 1,637 APIs, is called the \emph{complex APIs}. It can only be fully tested by input generators that support the creation of complex objects, which is in our case Chizpurfle{} in black box (BB) and evolutionary (Evo) mode.
The seven different setups are constructed by combining the fuzzer variants \afl~(BB), \afl~(Evo), \randfuzz, Chizpurfle (BB) and Chizpurfle (Evo) with the possible API groups they support.

\noindent
\paragraph{Performance.}
\label{discussion:eval:performance}

We conducted two performance evaluations:
First, we measure the raw performance of all fuzzers by computing the time between executions for generated inputs to evaluate their trade-off between more complex generation heuristics and sheer throughput of inputs.
Second, we use the fuzzers to measure the overhead introduced by our instrumentation modules for coverage tracking.

Table~\ref{tab:Performance} depicts the results of our measurements.
The first observation is \afl's low number of executions per second (between 30-150 as computed from the execution times) compared to the over 500 suggested by the \afl{} manual~\cite{bufferdensity}, which is expected because of the additional overhead added for the IPC round-trip and the execution of harness code. 
The second insight is that, as expected, \randfuzz{} is the fastest fuzzer because generating purely random values is magnitudes faster than the complex input generation logic of \afl. 

Third, as expected, the black box variant of Chizpurfle is faster than the evolutionary variant. While we would have expected similar results for \afl, we found large outliers in the measurements for the black box variant of \afl{} (with and without \artist~activated) that can explain the unexpectedly high execution time.  

The fourth information conveyed by the table is the acceptable overhead imposed by our instrumentation for the fuzzers.
Table~\ref{tab:Performance} indicates that our coverage tracking instrumentation is magnitudes faster than the Chizpurfle \frida{} module's reported 13-fold overhead.
However, we take this comparison with a grain of salt because, first, we could not evaluate \frida{} ourselves on our platform, and second, it was not explicitly stated in the papers~\cite{chizpurfle1,chizpurfle2} whether the overhead was an end-to-end measurement.

\begin{table}
	\caption{Performance measurements for time between executions in milliseconds, averaged over APIs using mean and median respectively.}
	\label{tab:Performance}
	\resizebox{\columnwidth}{!}{
		\setlength{\tabcolsep}{5pt}
		\definecolor{mygray}{gray}{0.85}
		\definecolor{myblue}{rgb}{0.74, 0.83, 0.9}
		\centering
		\begin{tabular}{lcccc}
			\toprule
			
			\multirow{2}{*}{Input Generator}& \multicolumn{2}{c}{Per-API Means} & \multicolumn{2}{c}{Per-API Medians}\\
			& Mean & Median & Mean & Median  \\
			
			\midrule
			\multicolumn{5}{c}{Complex APIs}	\\
			\midrule
			
			Chizpurfle Evo				& 387.97	& 22.95	& 374.24	& 20.88	\\
			Chizpurfle BB				& 99.15		& 8.88	& 101.61	& 7.68	\\
			Chizpurfle BB (no \artist)	& 75.92 	& 9.05	& 60.10		& 7.79	\\

			\midrule
			\multicolumn{5}{c}{Primitive APIs}	\\
			\midrule
			
			Chizpurfle Evo				& 170.47	& 23.30	& 166.99	& 20.98	\\
			Chizpurfle BB				&  15.57	& 8.74	& 13.98		& 7.52	\\
			Chizpurfle BB (no \artist)	& 15.69		& 8.84	& 13.96		& 7.62	\\
			
			\midrule
			
			AFL Evo						& 9.64		& 7.81	& 8.33		& 7.44	\\
			AFL BB						& 32.36		& 12.10	& 7.54		& 7.05	\\
			AFL BB (no \artist)			& 21.51		& 8.35	& 7.21		& 6.63	\\
			
			\midrule
			
			RandFuzz					& 0.42		& 0.08	& 0.34		& 0.07	\\
			RandFuzz (no \artist)		& 0.36		& 0.09	& 0.26		& 0.08	\\
			
			\bottomrule
		\end{tabular}
	}
	
\end{table}

\noindent
\paragraph{Coverage.}
\label{discussion:eval:coverage}
Table~\ref{tab:fuzzers} depicts the results of our evaluation that compares different input generators running in the same environment with identical time budget.
\begin{table}[h]
	\caption{Measured coverage for different input generators.}
	\label{tab:fuzzers}
	\resizebox{\columnwidth}{!}{
	
	\tiny
	\begin{tabular}{lcc}
		\toprule
		Input Generator					& coverage mean	& coverage median \\
		\midrule
		\multicolumn{3}{c}{Complex APIs} \\
		\midrule
		Chizpurfle Evo					& 28.31\% & 21.66\%	\\
		Chizpurfle BB					& 28.53\% & 21.88\%	\\
		\toprule
		\multicolumn{3}{c}{Primitive APIs} \\
		\midrule
		Chizpurfle Evo					& 36.63\% & 36.36\%	\\
		Chizpurfle BB					& 35.65\% & 33.33\%	\\
		AFL Evo							& 34.61\% & 33.33\% \\
		AFL BB							& 34.50\% & 29.73\%	\\
		RandFuzz						& 33.08\% & 29.79\% \\
	
		\bottomrule
	\end{tabular}
	}

\end{table}
In other domains, coverage-guided greybox fuzzers are commonly known to generate higher coverage than their black box equivalents so we want to evaluate whether this holds in our domain as well. The surprising result of our experiments is that this does \textit{not} seem to hold. As evident from Table~\ref{tab:fuzzers}, both setups that utilize evolutionary fuzzers, \afl{} (Evo) and Chizpurfle (Evo), fail to provide substantial benefits over their black box equivalents and the pure random \randfuzz{} in terms of achieved coverage during the testing. \textbf{This finding again reinforces that thorough analysis is needed to make principled decisions about the design of dynamic analyses for Android.}

\subsubsection{Bugs \& Crashes}
\label{usecases:vuln:results}

We also evaluated whether the  fuzzers were able to uncover bugs similar to what related work targeted. Our results show that, in contrast to related work that created specialized toolchains to aim at particular kinds of flaws, \textbf{integrating available fuzzers with our common platform already suffices to find instances of the same kinds without the need for specialization.} 
In the following, we will discuss the different categories of flaws that were detected by our system and relate them to those from related work.

\noindent
\textbf{Exception-based Crashes (UncaughtException).}
Similar to ExHunter, our system was able to find multiple \textsf{UncaughtException} flaws, which crash the systemserver and force a soft-reboot.
However, our system found these flaws without requiring any kind of specialization and---because it is built on top of \troop---goes the extra mile of confirming and reproducing it automatically.
The most surprising results of our evaluation, however, are three vulnerable APIs
that lead to crashes of the SystemUI app. Given the focus of our analysis, we mainly implemented custom logic for catching exceptions and native crashes in system services. However, because many problems are not known beforehand, we also created commands for our result analyzer that heuristically check the generated test artifacts for other interesting behavior by, e.g., collecting crash notifications from logcat, which led to the detection of these three SystemUI crashes. 
This discovery exemplifies how a more modular approach can lead to new insights, even beyond what was originally targeted.

\noindent
\textbf{Systemserver Freeze (ASV) + Bootloop.}
While ASVHunter was particularly designed to locate \textsf{Android Stroke Vulnerabilities (ASVs)}, our prototype was able to uncover an instance of this flaw in a generic way for the \TT{setOverscan} API in the \TT{WindowManagerService} (see Appendix~\ref{Apx:window:15}). Furthermore, the flaw we discovered not only reboots the systemserver once but permanently soft-bricks the whole device, which none of the existing works claim to have found.

\noindent
\textbf{Resource Exhaustion.}
We detected the vulnerable API \TT{getSharedAccountsAsUser} in the \TT{AccountManagerService} that, when hammered with fake user IDs for a period of less than a minute, opens a large number of SQLite database connections until the  file descriptor limit of 1024 is reached and the system crashes. Judging from the limited information that is available about related work, none of the proposed approaches seems to have found an instance of this flaw category.%

\noindent
\textbf{Native Crashes.}
During our evaluation, we surfaced a set of APIs that immediately forward all inputs they receive to their corresponding native counterparts via JNI. We then often saw a native systemserver crash generated by a tool called \component{CheckJNI}~\cite{checkjni} that prevents, e.g., dereferencing null pointers provided as inputs where complex Java objects are expected. While it is straightforward to misuse these as denial-of-service attacks for crashing the systemserver, we believe that these natively-implemented methods expose a completely uncharted attack surface that might lead to more serious vulnerabilities (e.g., memory corruption flaws), which we discuss in more detail in Section~\ref{CaseStudy:further}.

\noindent
\textbf{Exploitability.}
The \privflaws{} detected flaws require permissions that cannot be obtained by regular apps, %
thus our attacker model not only considers malicious third-party apps but in addition misused vendor apps. Prior research has shown that, in the past, vendors decreased platform security in their ROMs by, e.g., accidentally exposing a system shell \cite{samsungshell}, lowering protection levels of system permissions to dangerous or normal~\cite{harvesting}, or introducing confused deputy components that expose their privileges~\cite{vendorcustom}. Additionally, Xing et al.~\cite{permupgrade} reported a so-called permission upgrade attack where apps could elevate custom permissions to system permissions due to name clashes. These scenarios illustrate that this type of flaw can still be triggered by regular apps from the Google Play store in situations where the AOSP code is modified by vendors in an insecure way.

\subsubsection{Towards Principled Vulnerability \& Bug Hunting}

By exploiting the platform's modularity, we were able to compare different input generation strategies in terms of their performance, findings and generated coverage already. Our results provides valuable insights for future iterations of bug hunting dynamic analyses, such as to not focus on a single flaw category alone or to re-use existing building blocks. Furthermore, the modular integration provides additional functionality, such as automated post-analysis of results (e.g., auto-verification or identification of novel bug classes) for free.
There are many more aspects to be considered in the future, such as whether completely different input generators, like concolic executors, provide substantial benefits over fuzzers. 
However, our case study successfully highlights how a more structured approach allows to, for the first time in the field of systemserver dynamic analysis,
compare strategies from related work in a common testbed.

\subsection{Permission Mapping}

Our second case study focuses on dynamically creating a mapping of systemserver APIs to their sets of enforced permissions, which Stowaway implemented by instrumenting the systemserver to log permission checks and using the Randoop~\cite{randoop} test generator to create inputs. In this case study, we investigate alternative strategies for dynamic permission mapping by using existing fuzzers and comparing the results to state-of-the-art permission maps.

\subsubsection{Design}

The design is driven by the idea of re-instantiating the approach of Stowaway by exploiting the capabilities of our platform. We explore alternative input generation strategies by using the built-in fuzzers to test single APIs for a fixed amount of time. 
Instead of hooking the permission enforcement points, which were manually picked and modified in the original work, we automatically parse the exceptions thrown back to the fuzzer for enforced permissions. The number of failed permission checks is increased by running the fuzzer from within the context of a zero-permission app. 

\subsubsection{Comparative Evaluation}

\begin{table}
	\caption{Comparing the permissions mapped by different fuzzers.}
	\label{tab:permfuzzers}
	\resizebox{\columnwidth}{!}{
		\begin{tabular}{l|cc|cc}
			\toprule
			& \multicolumn{2}{c|}{Axplorer~\cite{axplorer}}	 & \multicolumn{2}{c}{Arcade~\cite{arcade}} 	\\
								& New			& 	Confirmed	& New			& Confirmed		\\
			\midrule
			& \multicolumn{4}{c}{Complex APIs}			\\
			\midrule
			Chizpurfle Evo 		& 132 			& 515 (42.08\%) & 287 			& 360 (41.67\%)\\
			Chizpurfle BB 		& 144 			& 557 (45.51\%)	& 307 			& 394 (45.60\%) \\
			\midrule
			& \multicolumn{4}{c}{Primitive APIs}			\\
			\midrule
			Chizpurfle Evo 		& 52 			& 206 (64.78\%)	& 110			& 148 (65.78\%)\\
			Chizpurfle BB 		& 56 			& 225 (70.75\%)	& 116 			& 165 (73.33\%) \\
			AFL Evo				& 50			& 229 (72.01\%) & 113			& 166 (73.78\%)	\\
			AFL BB				& 50			& 230 (72.32\%) & 112			& 168 (74.67\%)	\\
			RandFuzz			& 56			& 230 (72.33\%)	& 118			& 168 (74.67\%)	\\
			\bottomrule	
		\end{tabular}
	}
	
\end{table}

Since Stowaway is not available for newer Android versions, we have to compare our results to the Axplorer~\cite{axplorer} and Arcade~\cite{arcade} state-of-the-art static analysis permission mappings. As an example of how a common platform can help evaluate different design decisions, we ran the permission mapping experiment in all available fuzzer configurations. 

\begin{figure}
	\centering
	\includegraphics[width=0.9\columnwidth]{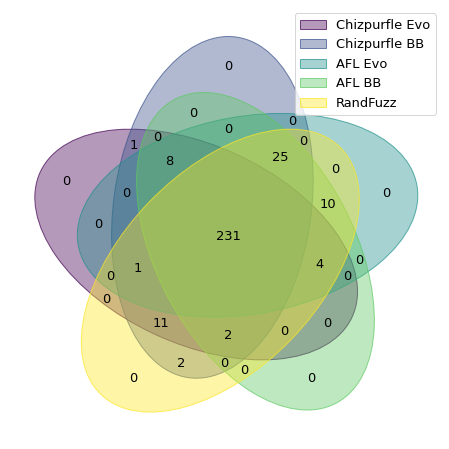}
	\caption{Overlap of mapped permissions for primitive APIs.}
	\label{fig:venn}
\end{figure}

\noindent
Table~\ref{tab:permfuzzers} depicts the results of our evaluation. While our prototype is already able to uncover new permission mappings missed by related work, it can not confirm all previous results. 
Figure~\ref{fig:venn} allows to further explore how the results relate to each other. 
Surprisingly, the fuzzers produce similar results despite their different input generation strategies (i.e., coverage-driven versus black box) and levels of sophistication. 
In our evaluation, coverage feedback as employed by Chizpurfle (Evo) and \afl{} (Evo) did \textit{not} lead to consistently superior results as evident from Table~\ref{tab:permfuzzers} and Figure~\ref{fig:venn}, which might be a distinctive factor of this particular problem domain. 
These findings are indicators that a principled approach to comparing and evaluating dynamic analyses down to their building blocks is required to challenge common beliefs and identify optimal strategies. 

\subsubsection{Towards Principled Permission Mapping}

Similarly to the first case study, instantiating this use case on top of our platform not only allows to utilize, e.g., the improved scalability and auto-verification,
but in particular, going forward, 
supports researchers to investigate and immediately compare different strategies for their analyses. %

\begin{punchline}
Our case studies show how a principled approach implemented by utilizing a common platform will allow future research to compare with and build upon related work, which allows us to effectively \textit{measure progress} in this area when evaluating novel approaches. 
\end{punchline}

\subsection{Further Use Cases}
\label{CaseStudy:further}

Besides the two use cases we implemented as case studies, there is a large set of possible dynamic analyses that can benefit from being instantiated on top of our platform.  

\noindent
First, in order to cope with the strong fragmentation and increased risk profile of the Android ecosystem due to custom vendor patches~\cite{samsungshell,vendorcustom,harvesting}, differential analysis can be used to analyze multiple ROMs and compare their observed behavior. 
Second, explicitly targeting the native code libraries used by the systemserver via JNI is another candidate that is yet to be explored by contemporary literature. While those libraries are only reachable indirectly through systemserver APIs, they are subject to more dangerous flaw classes such as memory corruption vulnerabilities.   
Third, similar to the approach of ASVHunter~\cite{asv}, our platform can also be used for testing static analysis findings in order to confirm or refute them, or even analyze targets beyond the middleware, such as apps, internal system libraries, or even the kernel.

\section{Discussion}
\label{Discussion}
We discuss Android version dependence, our disclosure process, and potential future extensions.%

\subsection{Android Version Support}
\label{VersionSupport}

We decided to start our analysis efforts with the Android Open Source Project (AOSP), Android 7.1 in particular, to make our findings immediately applicable to a large set of devices in the wild. 
However, using alternative Android versions instead (e.g., other major releases, custom ROMs) is just a matter of rooting the ROM and using the corresponding \artist{} version.
Since we strive for generality, the framework could also be applied to non-open source instantiations, such as ROMs from well-established Android vendors, like Samsung or Huawei, to reproduce results from previous works such as Chizpurfle that explicitly aimed at testing vendor services. There are two possibilities to achieve this: First, assuming that \artist{} supports instrumenting the systemserver without replacing Android's default compiler (this is work in progress according to~\cite{artistsystemissue}), our system could be applied to actual devices running non-open source ROMs. In particular, mapping the systemserver APIs once, preparing target devices by installing \artist{}, and changing the workers to additionally trigger the instrumentation modules should already suffice to make our infrastructure compatible with any closed source ROM.
Second, our framework is open source and thus can also be deployed on-premise by the vendor responsible for the ROM within their own infrastructure.

\subsection{Responsible Disclosure}
\label{Disclosure}

We responsibly disclosed all uncovered bugs to Google. While earlier works received CVEs for their findings, according to the wayback machine~\cite{wayback}, Google in the meanwhile changed the guidelines of their vulnerability disclosure program to explicitly remove application-triggered DoS attacks on the systemserver from the scope of the bug bounty program. In particular, related work, if published today, would also not receive CVEs for their findings.%

\subsection{Future Extensions}
We can exploit the platform's modularity to integrate more approaches from related work in the future to make them not only available to all existing analyses built on top but also bring more complex dynamic analyses within reach.%

\subsubsection{Transformational Fuzzing}
\label{TransformationalFuzzing}
Transformational fuzzing, as applied in T-Fuzz~\cite{tfuzz}, proposes to uncover new paths that are typically shielded by sanity checks by testing a transformed program to efficiently find candidates via fuzzing and verifying candidates via symbolic execution. Implementing this idea within our platform could tremendously boost the achieved coverage for all currently integrated input generators because it circumvents many of the checks that are hard to avoid by regular fuzzers, such as privilege checks or conditions on the environment. This approach is well suited for dynamic analyses that require a \textit{deep} analysis of the code.

\subsubsection{Test Case Minification}
\label{Minification}

Test case minification (also test case reduction, delta debugging) is a known technique~\cite{DeltaDebugging} in the software testing community to produce minimal and hence more efficient versions of test cases without reducing their utility. Similarly to how we implemented the automated verification of candidate inputs, our platform can incorporate standard techniques from this field to ensure the reproduction test cases are as small as possible with respect to size and runtime. Having a shared platform thereby allows to immediately apply this to all the different analyses built on top. 

\subsubsection{Alternative Input Generators}
\label{alternativeinput}

While we first focused on the integration of fuzzers into our platform, alternative input generators can be implemented as well. 
In particular, it would be interesting to combine our platform with frameworks such as angr~\cite{angr} or DeepState~\cite{deepstate} that already implement multiple input generation strategies, i.e., fuzzing \textit{and} concolic execution, in a unified framework. 
This would allow dynamic analyses to pick their strategy according to the needs of their concrete use case.

\subsection{Reproducibility}
\label{reproducibility}

To allow for open collaboration on this topic and independent evaluation of our tools and results, we are open sourcing our full platform along with documentation and setup scripts. Appendix~\ref{Apx:Tools} gives an overview of the 18 repositories.

\section{Conclusion}
\label{Conclusion}
In this work, we systematically analyzed common requirements and challenges when dynamically analyzing Android's crucial systemserver component. 
Based on our observations, we implement a common platform as a first step towards more generic solutions that allow to integrate, compare, and enhance existing and novel approaches in this area. 
We conducted two case studies that instantiate common analyses from related work where we utilized our platform's modularity to evaluate and discuss different implementation strategies and design decisions.
We hope that, going forward, our platform allows for a more principled way of approaching new interesting research ideas in the area of dynamically analyzing Android's middleware.

{\small \bibliographystyle{acm}
\bibliography{references}}

\begin{thebibliography}{10}

\bibitem{marketshare}
{Android Worldwide Market Share}.
\newblock \url{http://gs.statcounter.com/os-market-share/mobile/worldwide}.
\newblock Last visited: 27/05/2020.

\bibitem{aospbuild}
android/platform/build/nougat-release.
\newblock
  \url{https://android.googlesource.com/platform/build/+/nougat-release}.
\newblock Last visited: 27/05/2020.

\bibitem{artistgithub}
Artist - the android runtime instrumentation and security toolkit.
\newblock \url{https://github.com/Project-ARTist/ARTist}.
\newblock Last visited: 27/05/2020.

\bibitem{artistsystemissue}
{ArtistGui - Issue: System instrumentation support}.
\newblock \url{https://github.com/Project-ARTist/ArtistGui/issues/85}.
\newblock Last visited: 27/05/2020.

\bibitem{codelib}
Codelib.
\newblock \url{https://github.com/Project-ARTist/template-codelib}.
\newblock Last visited: 27/05/2020.

\bibitem{Flask}
Flask.
\newblock \url{http://flask.pocoo.org/}.
\newblock Last visited: 27/05/2020.

\bibitem{monkeytroop}
{Monkey Troop - ARTist's evaluation tool}.
\newblock \url{https://github.com/Project-ARTist/monkey-troop}.
\newblock Last visited: 27/05/2020.

\bibitem{chizpurflegithub}
{The Fantastic Beasts Framework for the Android OS}.
\newblock \url{https://github.com/dessertlab/fantastic\_beasts}.
\newblock Last visited: 27/05/2020.

\bibitem{arcade}
{\sc Aafer, Y., Tao, G., Huang, J., Zhang, X., and Li, N.}
\newblock {Precise Android API Protection Mapping Derivation and Reasoning}.
\newblock In {\em ACM CCS'18}.

\bibitem{harvesting}
{\sc Aafer, Y., Zhang, X., and Du, W.}
\newblock {Harvesting Inconsistent Security Configurations in Custom Android
  ROMs via Differential Analysis.}
\newblock In {\em USENIX Sec'16}.

\bibitem{asan}
{\sc {AOSP}}.
\newblock {AdressSanitizer}.

\bibitem{aospservice}
{\sc AOSP}.
\newblock {android/platform/frameworks/native/nougat-release/./cmds/service}.
\newblock
  \url{https://android.googlesource.com/platform/frameworks/native/+/refs/heads/nougat-release/cmds/service/}.
\newblock Last visited: 27/05/2020.

\bibitem{aosp}
{\sc AOSP}.
\newblock {The Android Open Source Project}.
\newblock \url{https://source.android.com/}.
\newblock Last visited: 27/05/2020.

\bibitem{pscout}
{\sc Au, K. W.~Y., Zhou, Y.~F., Huang, Z., and Lie, D.}
\newblock Pscout: analyzing the android permission specification.
\newblock In {\em ACM CCS'12}.

\bibitem{hongghof}
{\sc Authors, V.}
\newblock {GitHub: HonggFuzz - Trophies}.
\newblock \url{https://github.com/google/honggfuzz#trophies}.
\newblock Last visited: 27/05/2020.

\bibitem{randoop}
{\sc Authors, V.}
\newblock {Randoop - Automatic unit test generation for Java}.
\newblock \url{https://randoop.github.io/randoop/}.
\newblock Last visited: 27/05/2020.

\bibitem{axplorer}
{\sc Backes, M., Bugiel, S., Derr, E., McDaniel, P., Octeau, D., and
  Weisgerber, S.}
\newblock On demystifying the android application framework: Re-visiting
  android permission specification analysis.
\newblock In {\em USENIX SEC'16}.

\bibitem{artist}
{\sc Backes, M., Bugiel, S., Schranz, O., von Styp-Rekowsky, P., and
  Weisgerber, S.}
\newblock {ARTist: The Android Runtime Instrumentation and Security Toolkit}.
\newblock In {\em IEEE EuroS\&P'17}.

\bibitem{directedfuzzing}
{\sc B\"{o}hme, M., Pham, V.-T., Nguyen, M.-D., and Roychoudhury, A.}
\newblock Directed greybox fuzzing.
\newblock In {\em ACM CCS '17}.

\bibitem{buzzer}
{\sc Cao, C., Gao, N., Liu, P., and Xiang, J.}
\newblock {Towards Analyzing the Input Validation Vulnerabilities Associated
  with Android System Services}.
\newblock In {\em ACM ACSAC'15}.

\bibitem{angora}
{\sc {Chen}, P., and {Chen}, H.}
\newblock Angora: Efficient fuzzing by principled search.
\newblock In {\em IEEE S\&P'18}.

\bibitem{chizpurfle2}
{\sc Cotroneo, D., Iannillo, A.~K., and Natella, R.}
\newblock Evolutionary fuzzing of {Android} {OS} vendor system services.
\newblock {\em Empirical Software Engineering\/}.

\bibitem{cgc}
{\sc (DARPA), D. D.~F.}
\newblock {Cyber Grand Challenge (CGC) (Archived) }.
\newblock \url{https://www.darpa.mil/program/cyber-grand-challenge}.
\newblock Last visited: 27/05/2020.

\bibitem{lava}
{\sc Dolan-Gavitt, B., Hulin, P., Kirda, E., Leek, T., Mambretti, A.,
  Robertson, W., Ulrich, F., and Whelan, R.}
\newblock Lava: Large-scale automated vulnerability addition.
\newblock In {\em IEEE S\&P'16}.

\bibitem{androidafl}
{\sc ele7enxxh (GitHub~user)}.
\newblock {Fuzzing Android program with american fuzzy lop (AFL)}.
\newblock \url{https://github.com/ele7enxxh/android-afl}.
\newblock Last visited: 27/05/2020.

\bibitem{bindercracker}
{\sc Feng, H., and Shin, K.~G.}
\newblock {BinderCracker: Assessing the Robustness of Android System Services}.
\newblock {\em arXiv preprint arXiv:1604.06964\/} (2016).

\bibitem{collafl}
{\sc Gan, S., Zhang, C., Qin, X., Tu, X., Li, K., Pei, Z., and Chen, Z.}
\newblock Collafl: Path sensitive fuzzing.
\newblock In {\em IEEE S\&P'18}.

\bibitem{gong}
{\sc Gong, G.}
\newblock {Slides: Fuzzing Android System Services by Binder Call to Escalate
  Privilege}.
\newblock
  \url{https://www.blackhat.com/docs/us-15/materials/us-15-Gong-Fuzzing-Android-System-Services-By-Binder-Call-To-Escalate-Privilege.pdf}.
\newblock Last visited: 27/05/2020.

\bibitem{cts}
{\sc Google}.
\newblock {Compatibility Test Suite}.
\newblock \url{https://source.android.com/compatibility/cts}.
\newblock Last visited: 27/05/2020.

\bibitem{checkjni}
{\sc Google}.
\newblock {Debugging Android JNI with CheckJNI }.
\newblock
  \url{https://android-developers.googleblog.com/2011/07/debugging-android-jni-with-checkjni.html}.
\newblock Last visited: 27/05/2020.

\bibitem{ossfuzzhof}
{\sc Google}.
\newblock {GitHub: OSS-Fuzz - Trophies}.
\newblock \url{https://github.com/google/oss-fuzz#trophies}.
\newblock Last visited: 27/05/2020.

\bibitem{wayback}
{\sc {Google Bughunter University}}.
\newblock {Wayback Machine - Bugs with no security impact}.
\newblock
  \url{https://web.archive.org/web/20170110154209/https://sites.google.com/site/bughunteruniversity/android/invalid-bugs}.
\newblock Last visited: 27/05/2020.

\bibitem{acminer}
{\sc Gorski, S.~A., Andow, B., Nadkarni, A., Manandhar, S., Enck, W., Bodden,
  E., and Bartel, A.}
\newblock Acminer: Extraction and analysis of authorization checks in android's
  middleware.
\newblock In {\em CODASPY'19}.

\bibitem{arf}
{\sc Gorski, III, S.~A., and Enck, W.}
\newblock Arf: Identifying re-delegation vulnerabilities in android system
  services.
\newblock In {\em WiSec'19}.

\bibitem{he}
{\sc He, Q.}
\newblock {Slides: Hey your parcel looks bad - fuzzing and exploiting
  parcelization vulnerabilities in Android}.
\newblock
  \url{https://www.blackhat.com/docs/asia-16/materials/asia-16-He-Hey-Your-Parcel-Looks-Bad-Fuzzing-And-Exploiting-Parcelization-Vulnerabilities-In-Android-wp.pdf}.
\newblock Last visited: 27/05/2020.

\bibitem{asv}
{\sc Huang, H., Zhu, S., Chen, K., and Liu, P.}
\newblock From system services freezing to system server shutdown in android:
  All you need is a loop in an app.
\newblock In {\em ACM CCS'15}.

\bibitem{chizpurfle1}
{\sc Iannillo, A.~K., Natella, R., Cotroneo, D., and Nita-Rotaru, C.}
\newblock {Chizpurfle: A gray-box android fuzzer for vendor service
  customizations}.
\newblock In {\em IEEE ISSRE'17}.

\bibitem{wala}
{\sc IBM}.
\newblock {Wala - T.J. Watson Libraries for Analysis}.
\newblock \url{http://wala.sourceforge.net/wiki/index.php/Main\_Page}.
\newblock Last visited: 27/05/2020.

\bibitem{fuzzeval}
{\sc Klees, G., Ruef, A., Cooper, B., Wei, S., and Hicks, M.}
\newblock Evaluating fuzz testing.
\newblock In {\em ACM CCS'18}, ACM CCS '18.

\bibitem{undecidability}
{\sc Landi, W.}
\newblock Undecidability of static analysis.
\newblock {\em ACM LOPLAS'92\/}.

\bibitem{fans}
{\sc Liu, B., Zhang, C., Gong, G., Zeng, Y., Ruan, H., and Zhuge, J.}
\newblock {FANS}: Fuzzing android native system services via automated
  interface analysis.
\newblock In {\em USENIX SEC'20}.

\bibitem{binderdeath}
{\sc Lockwood, A.}
\newblock {Binder \& Death Recipients}.
\newblock
  \url{https://www.androiddesignpatterns.com/2013/08/binders-death-recipients.html}.
\newblock Last visited: 27/05/2020.

\bibitem{samsungshell}
{\sc Moulu, A.}
\newblock {Slides: Android OEM's applications (in)security and backdoors
  without permission}.
\newblock
  \url{https://www.sh4ka.fr/Android_OEM_applications_insecurity_and_backdoors_without_permission.pdf}.
\newblock Last visited: 27/05/2020.

\bibitem{deepstate}
{\sc of~Bits, T.}
\newblock {DeepState}.
\newblock \url{https://github.com/trailofbits/deepstate}.
\newblock Last visited: 16/11//2019.

\bibitem{tfuzz}
{\sc Peng, H., Shoshitaishvili, Y., and Payer, M.}
\newblock T-fuzz: fuzzing by program transformation.
\newblock In {\em IEEE S\&P'18}.

\bibitem{demystified}
{\sc Porter~Felt, A., Chin, E., Hanna, S., Song, D., and Wagner, D.}
\newblock Android permissions demystified.
\newblock In {\em ACM CCS'11}.

\bibitem{vuzzer}
{\sc Rawat, S., Jain, V., Kumar, A., Cojocar, L., Giuffrida, C., and Bos, H.}
\newblock Vuzzer: Application-aware evolutionary fuzzing.
\newblock In {\em NDSS'17}.

\bibitem{kAFL}
{\sc Schumilo, S., Aschermann, C., Gawlik, R., Schinzel, S., and Holz, T.}
\newblock kafl: Hardware-assisted feedback fuzzing for {OS} kernels.
\newblock In {\em USENIX SEC'17}.

\bibitem{frida}
{\sc Secure, N.}
\newblock {Frida - Dynamic instrumentation toolkit for developers,
  reverse-engineers, and security researchers.}
\newblock \url{https://frida.re}.
\newblock Last visited: 27/05/2020.

\bibitem{kratos}
{\sc Shao, Y., Ott, J., Chen, Q.~A., Qian, Z., and Mao, Z.~M.}
\newblock {Kratos: Discovering Inconsistent Security Policy Enforcement in the
  Android Framework}.
\newblock In {\em NDSS'16}.

\bibitem{angr}
{\sc Shoshitaishvili, Y., Wang, R., Salls, C., Stephens, N., Polino, M.,
  Dutcher, A., Grosen, J., Feng, S., Hauser, C., Kruegel, C., and Vigna, G.}
\newblock {SoK: (State of) The Art of War: Offensive Techniques in Binary
  Analysis}.
\newblock In {\em IEEE S\&P'16}.

\bibitem{soot}
{\sc Vall{\'e}e-Rai, R., Co, P., Gagnon, E., Hendren, L., Lam, P., and
  Sundaresan, V.}
\newblock Soot: A java bytecode optimization framework.
\newblock In {\em CASCON First Decade High Impact Papers'10}.

\bibitem{exhunter}
{\sc Wu, J., Liu, S., Ji, S., Yang, M., Luo, T., Wu, Y., and Wang, Y.}
\newblock {Exception Beyond Exception: Crashing Android System by Trapping in
  "uncaughtException"}.
\newblock In {\em ACM/IEEE ICSE'17}.

\bibitem{vendorcustom}
{\sc Wu, L., Grace, M., Zhou, Y., Wu, C., and Jiang, X.}
\newblock {The Impact of Vendor Customizations on Android Security}.
\newblock In {\em ACM CCS'13}.

\bibitem{permupgrade}
{\sc Xing, L., Pan, X., Wang, R., Yuan, K., and Wang, X.}
\newblock {Upgrading Your Android, Elevating My Malware: Privilege Escalation
  Through Mobile OS Updating}.
\newblock In {\em IEEE S\&P'14}.

\bibitem{bufferdensity}
{\sc Zalewski, M.}
\newblock {AFL - Understanding the status screen}.
\newblock \url{http://lcamtuf.coredump.cx/afl/status\_screen.txt}.
\newblock Last visited: 27/05/2020.

\bibitem{afl}
{\sc Zalewski, M.}
\newblock {american fuzzy lop}.
\newblock \url{http://lcamtuf.coredump.cx/afl/}.
\newblock Last visited: 27/05/2020.

\bibitem{aflhof}
{\sc Zalewski, M.}
\newblock {american fuzzy lop (2.52b) - The bug-o-rama trophy case}.
\newblock \url{http://lcamtuf.coredump.cx/afl/}.
\newblock Last visited: 27/05/2020.

\bibitem{DeltaDebugging}
{\sc Zeller, A.}
\newblock {Isolating Cause-effect Chains from Computer Programs}.
\newblock In {\em ACM SIGSOFT/SFE'02}.

\bibitem{drivers}
{\sc Zhou, X., Lee, Y., Zhang, N., Naveed, M., and Wang, X.}
\newblock The peril of fragmentation: Security hazards in android device driver
  customizations.
\newblock In {\em IEEE S\&P'14}.

\end{thebibliography}

\appendix
\onecolumn
\section*{Appendix}
\label{Appendix}
\section{Tools Overview (soon to be open sourced)}
\label{Apx:Tools}

\resizebox{\width}{!}{
	\setlength{\tabcolsep}{3pt}
	\definecolor{mygray}{gray}{0.85}
	\definecolor{myblue}{rgb}{0.74, 0.83, 0.9}
	\centering
	\begin{tabular}{ccccc}
		\toprule
		\textbf{Name} & \textbf{Description} & \textbf{Type} & \textbf{Languages} & \textbf{Forked} \\
		\midrule
		\multicolumn{5}{c}{Host-level} \\
		\midrule
		troop & Orchestrator & Framework & Python & monkey-troop\cite{monkeytroop}\\
		\rowcolor{mygray}
		db-backend & Task Backend & Server & Python & new \\
		avd-tool & Emulator Management & Script & Python & new \\
		\rowcolor{mygray}
		transaction-id-mapper & API to Binder IDs & Script & Python & new  \\
		crash-backend		& Crash \& Bug Collection + UI	& Server	& Python	& new	\\
		\midrule
		\multicolumn{5}{c}{On-Device} \\
		\midrule
		Chizpurfle	& Customized Fuzzer	& Java Executable	& Java	& chizpurfle~\cite{chizpurflegithub}	\\
		\rowcolor{mygray}
		android-afl & AFL for Android & Binary & C & andorid-afl\cite{androidafl} \\
		rand-fuzz & random fuzzer & Binary & C++ & new \\
		\rowcolor{mygray}
		libuplink & Ashmem FD Sharing & Service \& Lib & C++ & new \\
		Servitor & Test Harness & Binary & C++ & AOSP/Service\cite{aospservice} \\
		\rowcolor{mygray}
		FuzzWatch & Ashmem Visualization & Binary & C++ & new \\
		
		ARTist & Instrumentation & Framework & C++ & ARTist\cite{artistgithub} \\
		\rowcolor{mygray}
		art & ARTist Dependency & Runtime & C++ & AOSP/art\cite{aosp}\\
		
		Fuzzing Codelib & Injected Code Host & Java Library& Java & CodeLib\cite{codelib} \\
		\rowcolor{mygray}
		Replay App & Verification & App & Java & new \\
		\midrule
		\multicolumn{5}{c}{AOSP Build Support} \\
		\midrule
		fuzzing-scripts & Automation & Scripts & Bash & new \\
		\rowcolor{mygray}
		build & Extension & Makefile & Text & AOSP/build\cite{aospbuild} \\
		
		build\_fuzzing & Extension & Makefile & Text & new \\ 
		\bottomrule

	\end{tabular}
}

\section{Responsible Disclosure Report}
\label{Apx:Reports}

We responsibly disclosed the flaws described in  Section~\ref{usecases:vuln:results} to Google. In the following you can find one of those reports, here for the vulnerable \TT{WindowManagerService} API \TT{setOverscan}, as an example.

\label{Apx:window:15}
\bigskip
\subsection*{Title: Repeatedly calling \TT{WindowManagerService.setOverscan} soft-bricks the device (bootloop, broken SystemUI state)}
\bigskip
\noindent\textbf{TL;DR:}

The API \TT{com.android.server.wm.WindowManagerService.setOverscan(int,int,int,int,int)} can be abused by a privileged application/system component to send the device into an unstable state where the UI is unusable and there is nothing left but restarting the device. However, a restart triggers an infinite bootloop where the SystemUI cannot properly be instantiated (unusable for the user) and shortly afterwards \TT{Watchdog} kills the systemserver and everything repeats. We could only reclaim the device after deleting the userdata image via \TT{fastboot} (\textasciitilde~reset). The described vulnerability was automatically detected by our systemserver dynamic analysis system and manually confirmed afterwards on a Google Pixel running Android 7.1 (rooted, otherwise stock ROM). \\

\noindent\textbf{Fingerprint:} 

google/sailfish/sailfish:7.1.1/NOF27D/3757586:user/release-keys\\

\noindent\textbf{Description:}

The \TT{setOverscan} API can be trivially abused to deny the user access to their device by setting an overscan value that effectively moves all content outside the visible area. Additionally, if we keep on sending these requests in a loop, the window manager service lock is taken for too long so \TT{Watchdog} cannot execute its checker on the foreground thread, hence killing the systemserver. Now the system reboots but the SystemUI is broken, i.e. only the navigation bar is displayed but unusable. In particular, the \TT{sensorservice} that died earlier does not restart but the service manager needs to wait for it:

{\small
\begin{verbatim}
I ServiceManager: Waiting for service sensorservice...
E slim_daemon: [NDK] bindNDKSensors: Sensor server is unavailable.
I ServiceManager: Waiting for service sensorservice...
E slim_daemon: [NDK] bindNDKSensors: Sensor server is unavailable.
I ServiceManager: Waiting for service sensorservice...
E slim_daemon: [NDK] bindNDKSensors: Sensor server is unavailable.
E slim_daemon: [NDK] bindNDKSensors: Sensor server is unavailable.
I ServiceManager: Waiting for service sensorservice...
E slim_daemon: [NDK] bindNDKSensors: Sensor server is unavailable.
I ServiceManager: Waiting for service sensorservice...
E slim_daemon: [NDK] bindNDKSensors: Sensor server is unavailable.
I ServiceManager: Waiting for service sensorservice...
E slim_daemon: [NDK] bindNDKSensors: Sensor server is unavailable.
I ServiceManager: Waiting for service sensorservice...
E slim_daemon: [NDK] bindNDKSensors: Sensor server is unavailable.
I ServiceManager: Waiting for service sensorservice...
E slim_daemon: [NDK] bindNDKSensors: Sensor server is unavailable.
E slim_daemon: [NDK] bindNDKSensors: Sensor server is unavailable.
\end{verbatim}
}

From now on, the only thing the user can do is press the power button until the device restarts. However, now the device is kept in a boot loop because \TT{Watchdog} detects unresponsiveness:

{\small
\begin{verbatim}
W Watchdog: *** WATCHDOG KILLING SYSTEM PROCESS: Blocked in monitor com.android.server.wm.WindowManagerService
			on foreground thread 	(android.fg), Blocked in handler on ui thread (android.ui), Blocked in handler on 
			display thread (android.display)
W Watchdog: foreground thread stack trace:
W Watchdog:   at com.android.server.wm.WindowManagerService.monitor(WindowManagerService.java:11041)
W Watchdog:   at com.android.server.Watchdog$HandlerChecker.run(Watchdog.java:179)
W Watchdog:   at android.os.Handler.handleCallback(Handler.java:751)
W Watchdog:   at android.os.Handler.dispatchMessage(Handler.java:95)
W Watchdog:   at android.os.Looper.loop(Looper.java:154)
W Watchdog:   at android.os.HandlerThread.run(HandlerThread.java:61)
Watchdog:     at com.android.server.ServiceThread.run(ServiceThread.java:46)
Watchdog: ui thread stack trace:
W Watchdog:   at com.android.server.wm.WindowManagerService$LocalService.waitForAllWindowsDrawn
	(WindowManagerService.java:11663)
W Watchdog:   at com.android.server.policy.PhoneWindowManager.finishKeyguardDrawn
	(PhoneWindowManager.java:6385)
W Watchdog:   at com.android.server.policy.PhoneWindowManager.-wrap8(PhoneWindowManager.java)
W Watchdog:   at com.android.server.policy.PhoneWindowManager$PolicyHandler.handleMessage
	(PhoneWindowManager.java:773)
W Watchdog:   at android.os.Handler.dispatchMessage(Handler.java:102)
W Watchdog:   at android.os.Looper.loop(Looper.java:154)
W Watchdog:   at android.os.HandlerThread.run(HandlerThread.java:61)
W Watchdog:   at com.android.server.ServiceThread.run(ServiceThread.java:46)
W Watchdog: display thread stack trace:
W Watchdog:   at com.android.server.wm.WindowManagerService$H.handleMessage(WindowManagerService.java:8850)
W Watchdog:   at android.os.Handler.dispatchMessage(Handler.java:102)
W Watchdog:   at android.os.Looper.loop(Looper.java:154)
W Watchdog:   at android.os.HandlerThread.run(HandlerThread.java:61)
W Watchdog:   at com.android.server.ServiceThread.run(ServiceThread.java:46)
W Watchdog: *** GOODBYE!
\end{verbatim}
}

Full logcat dumps are attached. \\

\noindent\textbf{Proof-of-Concept:}

Our framework uses a version of the \TT{service} tool that forwards requests to the systemserver via Binder IPC. In order to confirm a potential issue, we simply replay the generated inputs using a script that repeatedly invokes \TT{service} with these inputs. The script and the inputs that trigger this behavior are attached.
For reproduction we currently use the adb shell b/c an app cannot obtain the required permission for changing the overscan, hence our attacker model is a misbehaving or buggy system app/component (potentially from a 3rd-party vendor). While this attacker model is strictly weaker than considering regular apps, e.g., from the play store, research has shown {[}1,2,3{]} that misbehaving vendor components including confused deputies existed in the past and that in some circumstances it is possible for regular apps to obtain system privileges. Note: We did not (yet) investigate whether there are eligible apps in AOSP or other ROMs that could be misused for the attack vector presented here.\\

\noindent\textbf{Author(s):} $\:$ $<$REDACTED$>$  \\

\noindent\textbf{Contact:}

Please feel free to contact us for any questions that might come up for this issue, we are happy to provide further information, receive feedback and engage in discussions.\\

\noindent\textbf{Attachments:}
\begin{itemize}
	\item script to reproduce findings via \TT{adb shell}
	\item input list that triggers the flaw
	\item logcat dumps: During attack \& after first reboot
\end{itemize}

\bigskip
\noindent
{[}1{]} https://www.sh4ka.fr/Android\_OEM\_applications\_insecurity\_and\_backdoors\_without\_permission.pdf \\
{[}2{} Paper: Harvesting Inconsistent Security Configurations in Custom Android ROMs via Differential Analysis \\
(https://www.usenix.org/system/files/conference/usenixsecurity16/sec16\_paper\_aafer.pdf)  \\
{[}3{]} Paper: The Impact of Vendor Customizations on Android Security \\
(http://citeseerx.ist.psu.edu/viewdoc/download?doi=10.1.1.687.360\&rep=rep1\&type=pdf)

\end{document}